\documentclass[aps,prb,a4paper,twocolumn,amsmath,amssymb,amsfonts,nofootinbib,floatfix,superscriptaddress]{revtex4-1}
\pdfoutput=1
\usepackage{graphicx} % include figure files
\usepackage[usenames, dvipsnames]{color}
\usepackage{dcolumn} % align table columns on decimal point
\usepackage{bm}
\usepackage{physics}
\usepackage{soul}
\usepackage{enumitem}
\usepackage{hyperref}
\usepackage{lipsum}
\usepackage[utf8]{inputenc}
\usepackage{natbib}
\usepackage{bookmark}
\hypersetup{
    citecolor=OliveGreen,
    colorlinks=true,
    urlcolor=blue
}

\definecolor{red}{rgb}{0.7,0.1,0.1}
\definecolor{Red}{rgb}{1,0,0}
\definecolor{green}{rgb}{0,0.8,0}
\definecolor{blue}{rgb}{0,0,0.8}
\definecolor{cautionred}{rgb}{1.0,0,0}
\definecolor{maroon}{rgb}{0.7,0,0}
\definecolor{ngreen}{rgb}{0.3,0.7,0.3}
\definecolor{golden}{rgb}{0.8,0.6,0.1}

\newcommand{\blk}{\color{black}}

\newcommand{\la}{\langle}
\newcommand{\ra}{\rangle}

\newcommand{\fref}[1]{Fig.~(\ref{#1})}
\newcommand{\eref}[1]{Eq.~(\ref{#1})}

\newcommand{\beq}{\begin{align}}
\newcommand{\eeq}{\end{align}}
\newcommand{\beqa}{\begin{align}}
\newcommand{\eeqa}{\end{align}}
\newcommand{\beqan}{\begin{align*}}
\newcommand{\eeqan}{\end{align*}}

\begin{document}

\title{%\instruction\\
  Finding the ground states of symmetric infinite-dimensional Hamiltonians: explicit constrained optimizations of tensor networks
}

\author{S.~N.~Saadatmand} \email{n.saadatmand@griffith.edu.au}
\affiliation{Centre for Quantum Dynamics, Griffith University, Nathan, QLD 4111, Australia.
}%
\date{\today}

\begin{abstract}
Understanding extreme non-locality in many-body quantum systems can help resolve questions in 
thermostatistics and laser physics.
The existence of symmetry selection rules for Hamiltonians with non-decaying terms on infinite-size lattices
can lead to finite energies per site, which deserves attention. Here, we present a  
tensor network approach to construct the ground 
states of nontrivial symmetric infinite-dimensional spin Hamiltonians based on constrained optimizations of their infinite
matrix product states description, which contains no truncation step, 
offers a very simple mathematical structure, 
and other minor advantages at the cost of slightly higher polynomial complexity in comparison to an existing method.    
More precisely speaking, our proposed algorithm is 
in part equivalent to the more generic and well-established solvers of infinite density-matrix renormalization-group and variational uniform matrix product states, which are, in principle, 
capable of accurately representing the
ground states of such infinite-range-interacting many-body systems. However, we employ
some mathematical simplifications that would allow for 
efficient brute-force optimizations of tensor-network matrices 
for the specific cases of highly-symmetric infinite-size infinite-range models. 
As a toy-model example, we showcase the effectiveness and explain some features of our method by finding the ground
state of the U(1)-symmetric infinite-dimensional antiferromagnetic $XX$ Heisenberg model. 
\end{abstract}

\maketitle

\section{Introduction}
Understanding the physics of many-body systems
exhibiting extreme non-local infinite-range interactions~\cite{Salerno1994,Bahn1999,Latora2001,Latora2002,Nobre2003,Baxter2007,Baumann2010,Fu2016,Chen2016,Flottat2017} (equal couplings between all subsystems with the coordination number $Z\!\rightarrow\!\infty$) in infinite dimensions is of great importance. Such Hamiltonians often appear in 
the thermodynamical studies of a wide range 
of contrived and engineered systems from classical Heisenberg ferromagnets (see in particular Ref.~\cite{Latora2001}) to quantum 
Dicke superradiance models (see in particular Ref.~\cite{Baumann2010}).
Yet, there exists only a single, and perhaps understudied, family of numerical methods capable of efficiently finding the phase diagrams of such Hamiltonians
for nontrivial scenarios as we discuss further below. 
%Existing ground state studies were limited to small finite-size lattices, some exactly solvable cases, 
%and not-so-precise mean field treatments. 

Let us first consider long-range Hamiltonians of the general form 
$\sum_{i>j,\frak{a}=x,y,z} \frac{J_{\frak{a}}}{r_{ij}^\alpha} \hat{S}^{\frak{a}}_i \hat{S}^{\frak{a}}_j$, where 
$r_{ij}$ denotes the distance between spins (or some other form of subsystems) 
$i$ and $j$, and $\alpha$ identifies the range of 
interactions. In the last four decades, such models have been consistently 
in the center of the attention due to exhibiting rich %and novel fundamentally important 
phase diagrams~\cite{Kardar1983,Kardar1983b,Rogan1997,Pluchino2004,Laflorencie2005,Baumann2010,Sandvik2010,Maik2012,Mottl2012,Gong2016,Humeniuk2016,Flottat2017,Saadatmand2018,Igloi2018a,Igloi2018,Koziol2019}, relevance to experimental cavity-mediated Bose-Einstein condensate~\cite{Baumann2010,Mottl2012,Klinder2015,Landig2016} %(BEC) 
or trapped ions~\cite{Britton2012,Bohnet2016} quantum simulators, and 
the emergence of nonextensive thermostatistics~\cite{Anteneodo1998,Salazar1999,Tamarit2000,Salazar2000,Latora2001,Latora2002,Pluchino2004,Mukamel2005,Apostolov2009,Campa2009,Bouchet2010} --- 
see~\cite{Campa2009}
for an extended review. The extreme case of 
infinite (global or all-to-all) range interactions corresponds to
$\alpha=0$ and, also, receives a great amount of 
attention as evident from Refs.~\cite{Salerno1994,Bahn1999,Latora2001,Latora2002,Nobre2003,Baxter2007,Baumann2010,Fu2016,Chen2016,Flottat2017}. One can think of this limit 
as the opposite case of highly local nearest-neighbor (NN) interactions having 
$\alpha\rightarrow\infty$, or the limit where the %physical dimension and 
lattice dimensionality and geometry become irrelevant. 

In general, the energy per site 
for an infinite-dimensional Hamiltonian with $\alpha\!=\!0$ non-decaying terms is diverging. Based on numerical experimentation on infinite-dimensional Heisenberg-type models (see~\cite{BakerInPrep} and also below) and intuition, we heuristically argue~\cite{McCullochComm} a finite energy-per-site exist when all higher-than-first moments or cumulants~\cite{Saadatmand2017_thesis} of the Hamiltonian operators are strictly zero. A general analytical proof is left for future works. (Notice for translation-invariant systems the energy-per-site can be always derived in terms of cumulants of the Hamiltonian operators~\cite{Saadatmand2017_thesis}. While the proof for the existence of a finite second cumulant/variance that leads to diverging energies-per-site is rather straightforward, it becomes cumbersome for the general case). As an example, for the U(1)-symmetric infinite-dimensional antiferromagnetic and ferromagnetic $XX$ Heisenberg models discussed below, the symmetry simply implies that the ground state must be in the $S^z\!=\!0$-symmetry-sector, which means the expectation value of $\hat{S}^z_{\rm total}$ (coinciding with the second and some higher-order cumulants of the Hamiltonian operators) is vanishing. 

Efficiently finding the phase diagrams and spectral degeneracy patterns of highly-symmetric 
infinite-range models can be regarded as an essential task of modern optics and thermostatistics due to
their appearance in some realistic and/or fundamentally important scenarios. In the following, we briefly
list few such examples. Most notably, quite recently it was realized that
the optical coherence of a continuous-beam laser can be regarded as an infinite-dimensional %symmetric 
effective Hamiltonian and has been studied~\cite{BakerInPrep} directly by employing the method we present below. 
Moreover, it is known that implementing out-of-equilibrium initial conditions in planar classical $N$-spin ferromagnets, which 
interact via an infinite-range potential and are collectively known as planar infinite-range Heisenberg mean field (HMF) model, lead to nonextensive thermodynamic~\cite{Latora2001,Latora2002,Pluchino2004} (i.e.~these systems would \emph{not} relax toward the conventional 
Boltzmann-Gibbs equilibrium distribution).
Importantly for such models, the case of $\alpha\!=\!0$ covers more than just fixed-coupling Hamiltonians: it is proven  
that the ground-state problem of HMF models on a $\frak{D}$-dimensional lattice having $0<\alpha<\frak{D}$ can be exactly reduced~\cite{Anteneodo1998,Tamarit2000,Campa2003} to an equivalent problem with $\alpha=0$. Another 
interesting example in the family of global-range-interacting systems are ultracold quantum gas systems fabricated to exhibit cavity-assisted infinite-range interactions. %Such systems have been now carefully considered: 
In one breakthrough work, a Dicke Hamiltonian was engineered and
a superradiant phase transition was observed experimentally~\cite{Baumann2010}. In another 
closely-related study, finite-size numerical simulations 
also elucidated the phase diagram of the two-dimensional infinite-range Bose-Hubbard model~\cite{Flottat2017}.

% classical ferromagnetic Ising~\cite{...}, 
% classical ferromagnetic Heisenberg XX~\cite{...} and XY~\cite{...}, 
% self-trapping bosonic~\cite{...} Hamiltonians, and Dicke's superradiance Hamiltonian~\cite{...}.
% An interesting phase transition phenomenon was observed recently in the lab~\cite{...} on the latter model.
% Many such infinite-range-interacting systems are, of course, known~\cite{...} to exhibit 
% thermostatistics distinct from conventional Boltzmann-Gibbs-Shannon equilibrium 
% behaviors in weakly-interacting gases; therefore, finding their . 

\subsection{Existing numerical methods for long-range infinite-size models}

Generally speaking, excluding some exactly-solvable cases (in particular, the Haldane-Shastry model~\cite{Haldane1988,Shastry1988} 
--- see also~\citep{Igloi2018}), 
finding the ground states
of \emph{translation-invariant} long-range Hamiltonians with arbitrary $\alpha$ is a challenging task
even in low dimensions and for unfrustrated systems. 
Highly-precise numerical methods, in principle, can tackle
such problems but have varied levels of applicability. 
In the forefront are some well-established variational tensor network approaches, which are based 
on matrix product states~\cite{AKLT,PerezGarcia2007,McCulloch2007,Schollwock2011,Saadatmand2017_thesis} (MPS) ansatz and the representation of $\frac{1}{r^\alpha}$ in terms of 
sum of a finite number of decaying exponentials (Pad\'{e} extrapolations), which are conventionally only 
applied to $\alpha > 1$ cases. One such powerful 
tensor-network solver is infinite density-matrix renormalization-group (iDMRG) 
method, based on the infinite MPS~\cite{McCulloch2008} (iMPS) and matrix product operator~\cite{McCulloch2007,McCulloch2008,Crosswhite2008b} (MPO) representations, 
which has been employed to scrutinize the ground states of a 
%one~\cite{...} and 
maximally frustrated two-dimensional long-range Heisenberg model~\cite{Saadatmand2018,Koziol2019}.
Notice another MPO-based algorithm was proposed~\cite{Crosswhite2008,Gong2016} prior to the two-dimensional iDMRG studies, where the authors investigated some
one-dimensional long-range Hamiltonians. However, such approaches are, in practice, equivalent to the iDMRG treatment~\cite{McCulloch2008,Saadatmand2018}.
Another generic tensor-network solver in this group
is variational uniform matrix product states (VUMPS)~\cite{ZaunerStauber2018} based on the MPS tangent space concept, which can be employed to find the phase diagrams of long-range 
%rapidly-decaying, but otherwise generic, 
Hamiltonians as efficient as (or more efficiently in some cases) the iDMRG method. 

In addition to tensor network approaches, exact diagonalization~\cite{Salerno1994,Sandvik2010,Fu2016} (ED) and 
quantum Monte Carlo~\cite{Bahn1999,Laflorencie2005,Maik2012,Humeniuk2016,Flottat2017} (QMC) simulations have been widely employed to study long-range models as well; however, indeed only for finite sizes (notice, as it is well-known, 
ED heavily suffers from 
the exponential growth in the Hilbert space size, while QMC faces the negative sign problem for such
calculations).
Mean-field theory approaches~\cite{Rogan1997,Baxter2007,Britton2012,Gong2016} could also provide valuable information on the phase diagrams of long-range-interacting systems, especially because some infinite-dimensional models have 
exact mean-field solutions~\cite{Baxter2007}, but often considered to have low validity due to the presence of inherently strong interactions for nontrivial cases.

Surprisingly, for the special case of infinite-range interactions, $\alpha\!=\!0$, highly-precise 
tensor network methods were never applied %and extended 
to infinite-dimensional (thermodynamic-limit) systems to our knowledge. However, these numerical routines can be 
prepared in a straightforward manner --- although possibly in different ways for different tensor 
network approaches --- to efficiently capture 
such cases as well, as we demonstrate below for the case of the iMPS.
(Notably, the slightly more-efficient iDMRG method can be also employed~\citep{McCulloch2008,McCullochComm} to construct ground states
comparable to what we report below; we have left presenting our iDMRG results on infinite-dimensional Hamiltonians
to a detailed future work.)
%for the tensor network case to be applicable
%to the general form of $J\sum_{i>j} \vektor{S}_i \cdot \vektor{S}_j$. 

\subsection{Results on global-range finite-size models}

First, it is noteworthy that the exact ground states 
are known for certain infinite-range models, most notably, the classical $N$-rotors HMF Hamiltonians~\cite{Anteneodo1998,Tamarit2000,Campa2003,Latora2001,Latora2002,Pluchino2004}. And we 
reiterate that insightful
mean-field~\cite{Baxter2007,Gong2016} and ED~\cite{Salerno1994,Fu2016} studies already exist for 
the case of \emph{finite-size} global-range
Hamiltonians. 
%(below, we also perform such systematic ED simulations to benchmark the results of our tensor network approach). 
More importantly, one can still perform the conventional variational finite-size DMRG~\cite{White1992,McCulloch2007,Fledderjohann2011,Hubig2015,Saadatmand2017_thesis} simulations to 
scrutinize phases of global-range models. %(as we also do below to benchmark our new results). 
In particular, the variational finite-size SU(2)-invariant MPS 
methods presented in Refs.~\cite{Ostlund1995,Fledderjohann2011} are in spirit similar to
the infinite-size approach we present below (in a sense that these references also directly 
diagonalizes and optimizes symmetric and sparse MPS matrices to converge to the ground state).
However, to scale up such finite-size tensor network results,
compensating for boundary effects, and
reaching to accurate ground-state energies would require very large bond dimensions and 
extreme system sizes.
Overall, the above finite-size algorithms are either imprecise or difficult 
to be applied to infinite-size systems, and currently, there seems to be 
a void in the existence of highly-precise numerical approaches, which are independent from renormalization routines and would target 
infinite-dimensional spin Hamiltonians. 
% Notably, after presenting main steps of our formalism below, it must becomes quickly clear that
% existing powerful iDMRG and VUMPS frameworks can be also partly altered (following the routines
% below) to find the phase diagram of the considered infinite-range model.

\subsection{Interior-point optimizations for infinite-dimensional models}

In this %relatively short 
paper, we demonstrate that the iMPS framework, independent from iDMRG, can be equipped with some mathematical 
simplifications to capture the ground states of infinite-dimensional models 
(by which we strictly mean both coordination number and lattice dimension diverge).
%In this paper, we introduce an efficient transfer-matrix-based tensor network 
%method for finding the ground states of, specifically,
Precisely speaking, our method is based on %`one-sweep' 
direct and highly-scalable constrained \emph{interior-point} optimizations 
of the parameters involved in the iMPS 
representation~\cite{BakerInPrep} of the physical states (see~Refs.~\cite{Byrd1999,Byrd2000,Waltz2006} for the 
interior-point optimization algorithm). 
%of infinite-size infinite-range Hamiltonians. 
Our independently-developed approach differs
from generic iDMRG and VUMPS %variational 
solvers due to the existence and absence of some %key 
features that makes it suitable only for capturing the physics of infinite-range models as detailed below.
While in principle, efficient iDMRG and VUMPS programs can be also prepared to represent infinite-range 
Hamiltonian terms,
here we are presenting a new brute-force algorithm potentially offering simpler 
implementation at the cost of slightly higher complexity. In our approach, it is needed to  
explicitly optimize a potentially large number of free parameters of the iMPS representation. 
However, at the same time, we provide an exact solution for the involved fixed-point equation, employ 
highly-scalable optimization steps, and most importantly,
discuss the built-in construction of Hamiltonian symmetries in this framework, which 
reduces the number of free parameters %to optimize for 
significantly. Overall, in this manner, we succeeded to provide a polynomial-cost tensor network algorithm, where the 
energy-per-site appears to rapidly converge to the true ground state
%scales polynomially with the inverse bond dimension 
as indicated below for an example. 
%(making our 
%approach still tractable).

We explain some major features and showcase the effectiveness of
our method by precisely finding the ground state for the working example of 
the U(1)-symmetric infinite-dimensional antiferromagnetic XX Heisenberg model. 
%(although, we are only 
%exploiting a U(1)-symmetric representation below due to pedagogical reasons). 
%-- note even with built-in symmetries this problem is not tractable
%to be solved using existing methods). 
We expect that the extension of our approach to other infinite-range models and Hamiltonian symmetries is straightforward.

The rest of the paper is organized as follows. In Sec.~II, we review some basic concepts of the iMPS description
and introduce our notation. The main expectation value of the energy per site for the infinite-range XX Heisenberg model
is derived in Sec.~III. The details of the constrained interior-point optimizations of the iMPS ansatz for this 
example are provided in Sec.~IV. Next, in the same section, we discuss the connections and some %key 
differences of this algorithm with the well-established tensor-network solvers of iDMRG and VUMPS. Finally, we benchmark the energies from our tensor network 
approach against an exact reference value %from fDMRG simulations (exploiting the same built-in symmetry) 
in Sec.~V, and end with a conclusion.

\section{The \lowercase{i}MPS representation}
In this section, we briefly review %and extend 
the iMPS representation of one-dimensional 
translation-invariant states. The iMPS ansatz is indeed a suitable choice for the representation of eigenstates of infinite-dimensional
Hamiltonians as it shall become clear below. We will only focus on the details that are particularly relevant 
to our goal here; for a full review 
of the iMPS formalism see~\cite{McCulloch2008,Schollwock2011,Saadatmand2017_thesis}.

\subsection{The essentials}
Generally speaking, the iMPS ansatz offers an approximate representation for translation-invariant
physical states. For 
\emph{one-site unit-cell} sizes, %(the extension to many-site unit-cells is straightforward through coarse-graining), 
this representation can be written as
\begin{align}
  \ket{\Psi_{\rm iMPS}}\!=\!\sum_{...,j_{-1},j_{0},j_{1},...} \text{Tr}\big( \cdots A^{[j_{-1}]} 
                           A^{[j_{0}]} A^{[j_1]} \cdots \big) \notag\\ 
                           \ket{...,j_{-1},j_{0},j_{1},...}~,
\label{eq:MPS-WF}
\end{align}
where $A^{[j]}$ denotes the usual $D \times D$ MPS $A$-matrices with $D$-dimensional virtual bonds and
$j=1,\cdots,d$ goes through the $d$-dimensional physical space of constituent particles. 
The representation is essentially exact for any quantum state when $D\rightarrow\infty$, 
therefore $D^{-1}$ can be considered as a precision control parameters --- the complexity
of tensor network algorithms often remain polynomial against $D$, and therefore, such ans\"{a}tze are
considered tractable, virtually exact, and can be handled on classical machines 
(note, however, the relevant energy errors can blow up exponentially against $D^{-1}$ for some tensor network
approaches making them inefficient classically).
%; our energy results below indicates this is not the case for the presented method).
There are remaining degrees of freedom in the above representation; therefore,
without loss of generality, we can assume that the orthonormality relation of 
$\sum_{j} A^{[j]}{}^\dagger A^{[j]} = I_D$ always holds, where $I_m$ denotes 
the $m \times m$ identity matrix. Furthermore, $A$-matrices must 
satisfy the fixed-point relation of $\sum_{j} A^{[j]} \rho A^{[j]}{}^\dagger = \rho$, where
$\rho$ is the diagonal right (reduced) density matrix -- more details below. In other words, we are assuming 
that the iMPS is already placed in the  
left-orthonormal/canonical form~\cite{Vidal2003}.

\subsection{Transfer matrix approach and the flattened space}

We intend to evaluate thermodynamic-limit expectation values (in particular the energy per site) using the 
well-established method of MPS transfer 
operators/matrices; see~\citep{McCulloch2008,Michel2010,Schollwock2011,Orus2014,Saadatmand2017_thesis,ZaunerStauber2018,BakerInPrep} for an introduction 
and useful graphical notations of MPS transfer operators.
Let $\mathcal{T}_{\hat{X}}$ denote a transfer operator equipped with 
the local physical operator $\hat{X}$. (Note that $\mathcal{T}$-matrices 
are, in fact, superoperators themselves acting on $D \times D$-size MPS operators.) 
Most significant in this superoperator family is the identity transfer operator, 
which will be shown as $\mathcal{T}_{\hat{I_4}} \equiv \mathcal{T}$.
The actions of $\mathcal{T}$ on two left- and right-hand-side MPS operators
can be then written as
\begin{align}
  \mathcal{T}(\hat{E})_{\rm left} &= \sum_j A^{[j]}{}^\dagger \hat{E} A^{[j]} \notag \\
  \mathcal{T}(\hat{F})_{\rm right} &= \sum_j A^{[j]} \hat{F} A^{[j]}{}^\dagger~.
\label{eq:T-L&R-operators0}
\end{align}
%%%

While working in this framework, it is more convenient to employ the so-called \emph{flattened space} notation (see for example Refs.~\cite{ZaunerStauber2018,BakerInPrep}). 
One can always reshape a $D \times D$-size operator into a flattened $D^2 \times 1$-dimensional vector 
form as $\hat{E}_{m,n}\rightarrow ( E |_{(m,n)}$ and $1 \times D^2$-dimensional 
vector form of $\hat{F}_{m,n}\rightarrow | F )_{(m,n)}$,
where $(m,n)$ stands for a collective index and $m,n=1,\cdots,D$.
In this flattened space, the transfer-type operators become large $D^2 \times D^2$ matrices and MPS operators are represented by $D^2$-size vectors.
Therefore, one can use a bra- and ket-like notation to write the left- and right-hand-side acting vectors in the flattened space language:
\begin{align}
  \big( ( E | \mathcal{T} \big)_{l l^\prime} &= \sum_{j,m,n} ( E |_{(m,n)} (A^{[j]}{}^\dagger)_{lm} A^{[j]}_{nl^\prime} \notag \\
  \big( \mathcal{T} | F ) \big)_{l l^\prime} &= \sum_{j,m,n} A^{[j]}_{lm} (A^{[j]}{}^\dagger)_{nl^\prime} | F )_{(m,n)}~.
\label{eq:T-L&R-operators}
\end{align}

In the flattened space, the transfer matrix can be constructed as
$\mathcal{T} = \sum_j A^{[j]}{}^* \otimes A^{[j]}$. We restrict ourselves 
to (perhaps physically more interesting) \emph{injective} $\mathcal{T}$-operators; therefore,
$\mathcal{T}$ has a unique pair of nonnegative left and right leading eigenvectors,
$\{|\lambda_1=1),(\lambda_1=1|\}$ and the 
spectral radius of $1$ (refer to the quantum version of the 
Perron-Frobenius theorem~\citep{Albeverio1978,Farenick1991} --- notice $\mathcal{T}$ is generally speaking
a non-hermitian matrix). In addition,
due to the orthonormality condition above, the left leading
eigenvector/eigenmatrix is the identity operator $\hat{I}_D \leftrightarrow (1| $,
i.e.~$(1| \mathcal{T} = (1| \lambda_1 = (1|$ in the flattened space language. 
%We further suppose that the eigenvalues of $\mathcal{T}$ are arranged as 
%$1=|\lambda_1| \geq  |\lambda_2| \geq \cdots \geq |\lambda_{D^2}|$.
Finally, due to the fixed-point equation above, the corresponding right eigenmatrix is the familiar 
reduced density matrix, ${\rho}\leftrightarrow |1)$, i.e.~$\mathcal{T} |1) = \lambda_1 |1) = |1)$.

If the spectrum or even leading eigenvalues of a well-converged (to the 
ground state of a physical model) iMPS transfer operator are known, all 
thermodynamic-limit expectation values can be found exactly or precisely---in particular, 
the second largest eigenvalue %$\lambda_2$, 
specifies the principal correlation length of the system and is typically
enough to estimate ground state expectation values (see \citep{Michel2010,Saadatmand2017_thesis} for details).
Note that the full diagonalization of the $\mathcal{T}$-matrix explicitly 
is rather a difficult numerical task; in general, 
it is a $D^2\!\times\!D^2$ non-sparse non-hermitian matrix. Instead, one can employ some mathematical simplifications 
to make the direct optimization of $\mathcal{T}$ significantly more 
efficient, which forms the essence of the current work. Here, we detail a
transfer operator approach that does \emph{not} require the direct calculation of the spectrum of $\mathcal{T}$ and is specifically useful to find the expectation values of symmetric infinite-range Hamiltonians.  
(However, note that our approach in this regard is comparable to 
subspace diagonalization of the relevant symmetry blocks of $\mathcal{T}$, employed in techniques like VUMPS, to efficiently find the required leading eigenvalues having an 
$O(D^3)$ cost.)
In a sense, our method 
%due to the 
%absence of conventional optimization algorithms involving many sweeps of the unit-cell, 
%as in iDMRG, for 
%infinite-range interactions (remember Pad\'{e} extrapolations are only valid for $\alpha>1$), 
%exploits some mathematical simplifications to directly
optimizes the element of the $\mathcal{T}$-matrix conditioned to the existence of some Hamiltonian rules and 
%works only by \emph{one} go but 
involves many interior-point optimization iterations as detailed further below.
%(there are no multiple sweeps on the unit-cell sites in our algorithm to converge the $A$-matrices to an accurate enough representation). 

\section{Writing down the energy per site for the representative example}

From this point onward, it is useful to 
present the remaining technical details using the %working 
toy-model example of
the spin-$1$ antiferromagnetic infinite-dimensional
$XX$ Heisenberg model in a zero field. However, the following formalism can be easily extended to other nontrivial, but highly-symmetric, 
infinite-range-interacting infinite-size systems. The Hamiltonian 
for the $XX$ model of our interest can be written as
\begin{align}\label{eq:XX-Ham}
  H_{XX} = J \sum_{i < j} (\hat{S}^+_i \hat{S}^-_j+\text{h.c.})
\end{align}
where $i$ and $j$ go over all spins and we set $J=1$ as the unit of energy
(notice the magnitude of $J$ has no physical importance for this model). Variants of
the XX model were previously carefully investigated due 
to their foundational importance and connections to 
some experiments; however, 
%Igl\'{o}i et al.~\citep{Igloi2018} investigates the phase diagram and out-of-equilibrium dynamics of a spin-1/2 NN ferromagnetic $XX$ Heisenberg model with competing infinite-range interactions and transverse field; 
% The model is exactly solvable and equivalent to the Bose-Hubbard Hamiltonian of cavity-mediated infinite-range-interacting 
% systems stabilized experimentally~\citep{Baumann2010,Mottl2012,Klinder2015,Landig2016} in the hard-core boson limit.
% The phase diagram of spin-$1$ one-dimensional NN Hamiltonian, itself, was recently studied~\citep{Malvezzi2016} by 
% employing fDMRG. Interestingly, the ground state of the NN model was indeed found to be an Haldane paramagnetic 
% phase. However, 
to our knowledge, the ground state of \eref{eq:XX-Ham} was \emph{not} constructed in the past and  
is relatively challenging to be found using conventional numerical methods in the thermodynamic limit. It is noteworthy that the ground state for the NN version of \eref{eq:XX-Ham} is a critical phase, which lives on the XY to Haldane phase transition point of the antiferromagnetic nearest-neighbor XYZ Heisenberg model~\cite{Kitazawa1996,Malvezzi2016}.   

\subsection{Exploiting Hamiltonian symmetries}

We start by looking for the Hamiltonian symmetries: the $A$-matrices that would 
represent the eigenstates of $H_{XX}$ have a highly 
reduced number of free parameters due to the presence of the 
Abelian U(1)-symmetry (note the Hamiltonian in \eref{eq:XX-Ham} commutes 
with the $\hat{S}^z_{\rm total}$-operator).
Importantly, we observe and confirmed numerically that due to the presence of the symmetry, the second and higher-order cumulants of the Hamiltonian 
operators remain zero by structure in the 
ground state symmetry sector, which results in a finite ground state energy per site in the thermodynamic limit.
Working with an irreducible iMPS representation, having 
built-in U(1)-symmetry, is indeed a very efficient way to find the ground state
of the model. 
%Therefore, we only work with the U(1)-symmetric version of $A$-matrices here, 
%which has a relatively simpler mathematical structure but more free parameters to optimize for.
We argue this U(1)-symmetric implementation is suitable for pedagogical reasons and
will prove the robustness of our scheme in finding the ground state 
in the case of choosing/realizing the symmetry in the model appropriately. 
(We reiterate that this built-in implementation of the symmetry can be 
extended to non-U(1) cases as well --- see for example Refs.~\citep{McCulloch2007,Saadatmand2017_thesis}.)

It is well-known~\citep{McCulloch2007,Saadatmand2017_thesis} that the U(1)-symmetry limits the number of 
elements in $A$-matrices \emph{allowed to be nonzero} and lead to a block diagonal structure 
in $\mathcal{T}$-operators that we exploit below. We arbitrarily choose 
the symmetry convention as $A^{[j]}_{m,n} \neq 0 
~\text{iff}~ m+j=n$, where $j$ corresponds to the $S^z$ quantum number. 
%of a spin site. 
Therefore, the three $D \times D$ iMPS $A$-matrices of a translation-invariant spin-1 system
can be shown as
\begin{align}
	A^{[-1]} = \left(
\begin{matrix}
0  & \cdots  & \cdots & \cdots & 0 \\
\bullet  & \ddots&& & \vdots \\
0  & \bullet & \ddots& &\vdots\\
\vdots  & \ddots & \ddots &\ddots & \vdots\\
0 & \cdots &  0 & \bullet & 0 \\
\end{matrix}
\right)
,
\end{align}
\begin{align}
A^{[0]} = \left(
\begin{matrix}
\bullet  & 0  & \cdots & \cdots & 0 \\
0  & \bullet& \ddots &  & \vdots \\
\vdots  & \ddots & \ddots &\ddots & \vdots\\
\vdots  &  & \ddots&\bullet & 0\\
0 & \cdots &  \cdots& 0 & \bullet\\
\end{matrix}
\right),
\end{align}
\begin{align}
A^{[1]} =  \left(
\begin{matrix}
0  & \bullet  & 0 & \cdots & 0 \\
\vdots  & \ddots&\bullet& \ddots & \vdots \\
\vdots  &  & \ddots & \ddots & 0\\
\vdots  &  & &\ddots & \bullet\\
0 & \cdots &  \cdots& \cdots & 0 \\
\end{matrix}
\right)~,
\label{eq:A-forms}
\end{align}
where bullets indicate the only elements allowed to be nonzero. In addition, we
choose all \emph{real-valued} $A$-matrices to represent the ground state due to the time-reversal 
symmetry of $H_{XX}$. Note also
the left-handed orthogonality condition 
% immediately reduces to
% %
% \begin{align}
%   (A_{m+1,m}^{[-1]})^2 + (A_{mm}^{[0]})^2
%   + (A_{m-1,m}^{[+1]})^2 = 1, %\quad\forall m~,
% \label{eq:ExplicitLeftOrth}
% \end{align}
% %
% which additionally 
implies that the absolute value of all $A$-matrices' elements 
are bounded from above by unity, $|A^{[j]}_{mn}| \leq 1~\forall \{j,m,n\}$. Furthermore, 
%we arbitrarily   
%choose to optimize for all real and nonnegative elements, $A^{[j]}_{mn} \geq 0 \forall \{j,m,n\}$,
%for simplicity of numerics.
The ground state must belong to the unit-cell $S^z\!=\!0$ symmetry sector, which further implies
\begin{align} \label{eq:Sz0-sector}
\la \hat{S}^z_{\rm total} \ra = 0~.
\end{align}
However, importantly, this condition is always automatically satisfied due to the structure $A$-matrices above. We still explicitly report this constraint below
for completeness (it may need to be enforced for other problems and basis choices). 

%The number of free parameters
%can be reduced even further as a result of the iMPS structure we have intentionally employed 
%in \eref{eq:MPS-WF}. 
Using the left-orthogonality and fixed-point equations it is straightforward to  
derive the following \emph{exact} recursive solution 
for the nonzero elements of right reduced density matrices 
(${ \rho}_{mm}\equiv{ \rho}_{m}$) of an iMPS of the form in \eref{eq:A-forms}:
\begin{align} \label{rhossAs}
{ \rho}_{m} = \left( \frac{A^{[-1]}_{m,m-1}}{A^{[+1]}_{m-1,m}} \right)^2 { \rho}_{m-1}~,~~0< m < D. 
\end{align}
In other words, if the $A$-matrices are known, the above will fully determine ${\rho}$ (or equivalently $|1)$ ---
the first diagonal element of the density matrix can be found by assuming a normalization for it). 
% In fact, upon numerical investigations, we have realized that by fixing this further $D-1$ free parameters in $\rho^{\rm ss}$ (which ties $A^{[0]}$ and $A^{[3]}$ via the above recurrence relation), the optimum $A$-matrices can be now found in a rather efficient manner using direct constrained optimizations.
% Notice there are $D-1$ parameters left to be optimized to reach an accurate representation of the ground state.
Notice \eref{rhossAs} is, in general, valid for all U(1)-symmetric cases.
To this end, there are overall $3D-2$ free parameters in 
%the ground state symmetry sector 
%($S^z_{\rm unit-cell}=0$), considering all 
the $A$-matrices 
and nonnegative $\rho$ to be optimized alongside strictly satisfying the constraints as we list below.

\subsection{Dealing with infinite sums}
Now consider the important quantity of the energy per site for the Hamiltonian of \eref{eq:XX-Ham},
which can be expressed as $\frak{e} = 2 \sum_{i>0} \la \hat{S}^+_0 \hat{S}^-_i+\text{h.c.} \ra$.
This can be written in the language of $\mathcal{T}$-operators discussed above
as follows
\begin{align}
  \frak{e} = 2\sum_{r=0}^\infty (1| \mathcal{T}_{S^+} \mathcal{T}^{r} 
    \mathcal{T}_{S^-} + \mathcal{T}_{S^-} \mathcal{T}^{r} 
    \mathcal{T}_{S^+} |1)~.
\label{eq:C-in-Tmatrices}
\end{align}
Most notably, our algorithm is based on writing the above form of a thermodynamic-limit expectation value 
in a reduced eigenvector space of $\mathcal{T}$; we intend to find a relevant inverse form of $\mathcal{T}$, and 
then perform highly-scalable constrained optimizations on its elements as we detail further below.  

It can be easily observed that the vector space of $\lambda_1=1$ has no contribution to the left-hand side term in 
\eref{eq:C-in-Tmatrices} (also happens because the ground state is constrained to the symmetry sector of $S^z=0$). Additionally, we are generally interested in using
a geometric-series-type relation to simplify that equation. Therefore, we project out this 
space from the set of eigenvectors
by defining the following projector:
\begin{align}
  \mathcal{Q} = I_{D^2} - |1)(1|~,
\label{eq:Q-projector}
\end{align}
which implies $\mathcal{T} = \mathcal{Q} \mathcal{T} \mathcal{Q} + |1)(1|$.  
Replacing $\mathcal{T}$ with this expression in \blk
\eref{eq:C-in-Tmatrices} leads to
\begin{align}
  \frak{e} = 2\sum_{r=0}^\infty \big\{ (1| \mathcal{T}_{S^+} (\mathcal{Q}\mathcal{T}\mathcal{Q})^r   \mathcal{T}_{S^-} |1) + (1| \mathcal{T}_{S^-} (\mathcal{Q}\mathcal{T}\mathcal{Q})^r   \mathcal{T}_{S^+} |1) \big\}~,
\label{eq:C-progress}
\end{align}
where we used $(1| \mathcal{T}_{S^+} |1)=(1| \mathcal{T}_{S^-} |1)=0$. 

Now, the superoperator $\mathcal{Q} \mathcal{T} \mathcal{Q}$ in the first term of \eref{eq:C-progress} has no unity eigenvalue. 
Therefore, the inverse of the object $I_{D^2} - \mathcal{Q} \mathcal{T} \mathcal{Q}$ will be well-defined. 
The infinite sums appearing in the first terms of Eq.~\eqref{eq:C-progress} can be replaced 
using geometric-series-type identities leading to
\begin{align}
  \frak{e} = 
%             &2\{(1| \mathcal{T}_{S^+} \cdot \text{inv}(I_{D^2} - 
%              \mathcal{Q}\mathcal{T}\mathcal{Q}) \cdot \mathcal{T}_{S^-} |1) \notag \\
%             &+ (1| \mathcal{T}_{S^-} \cdot \text{inv}(I_{D^2} - 
%              \mathcal{Q}\mathcal{T}\mathcal{Q}) \cdot \mathcal{T}_{S^+} |1)\} \notag \\
             2\{(S^+| \bar{\mathcal{T}}^{-1} |S^-)
            + (S^-| \bar{\mathcal{T}}^{-1} |S^+)\}~,
\label{eq:C-final}
\end{align}
where we have employed the shorthand notations of $(X| \equiv (1| \mathcal{T}_{X}$, 
$|X) \equiv \mathcal{T}_{X} |X)$, and $\bar{\mathcal{T}}\equiv I_{D^2} - \mathcal{Q}\mathcal{T}\mathcal{Q}$. 
Note that $\bar{\mathcal{T}}$-type matrices are often badly scaled and almost singular. One can efficiently estimate the above expression using either Moore-Penrose inverse or regularizing $\bar{\mathcal{T}}$
by adding a small $\epsilon I_{D^2}$ term and then proceed by standard discrete inversion methods for sparse matrices; one can also estimate the above by implicitly calculating $\bar{\mathcal{T}}^{-1} |X)$ terms using an iterative sparse Krylov-based solver. 
%which is what we have done for calculations below.
Equation \eqref{eq:C-final} is our main recipe to calculate the expectation values of thermodynamic-limit
energy-per-sites for infinite-range models, which we use for explicit numerical optimizations. Furthermore, this
can be easily extended to other thermodynamic-limit quantities of interest. 

\section{Constrained optimizations of the \lowercase{i}MPS}
To this end, for a given (possibly large) $D$-value, we 
require to efficiently find $A$-matrices elements that minimizes \eref{eq:C-final} subjected
to the following set of constraints for $3D-2$ free parameters:
\begin{align}
   \begin{cases}
     (1)~ \text{All the forms prescribed in \eref{eq:A-forms}}~, \\
%      (2)~ (A_{m+1,m}^{[-1]})^2 + (A_{mm}^{[0]})^2
%           + (A_{m-1,m}^{[+1]})^2 = 1, \\
     (2)~ \sum_{j} A^{[j]}{}^\dagger A^{[j]} = I_D~, \\
     (3)~ \la \hat{S}^z_{\rm total} \ra = 0~, \\     
     (4)~ { \rho}^{\rm ss}_{m} = \left( \frac{A^{[-1]}_{m,m-1}}{A^{[+1]}_{m-1,m}} 
           \right)^2 { \rho}^{\rm ss}_{m-1}~,~0< m < D~.
%     (5)~ A^{[1]} = \sqrt{I_D - A^{[0]\dagger} A^{[0]} - A^{[3]\dagger} A^{[3]}}~. \\ 
%     (6)~ \rho^{\rm ss} = \frac{1}{N_0(p)} \sum_{m=0}^{D-1} \sin^p( \pi \frac{m+1}{D+1} ) \ket{m} \bra{m} ~~\text{and} \\ 
%     ~~~~~\rho_{m} = \left( \frac{\kappa}{\sqrt{2} A^{[3]}_{m-1,m}} \right)^2 \rho_{m-1},~~m>1~. 
   \end{cases}
\label{eq:PoissonianCollectionAnsatz}
\end{align}
We argue (and demonstrate below for the example of $H_{XX}$) that %for relatively large $D$-values 
the ground state can be accurately found by minimization of the appropriately constructed cost function, in our case the expectation value
in \eref{eq:C-final}, constrained to \eref{eq:PoissonianCollectionAnsatz}, by employing
conventional highly-scalable constrained numerical optimization tools based on finite difference methods. 
We choose the interior-point optimization method as we found out that it
works increasingly well when adding 
more degrees of freedom to the set of $A$-matrices (i.e.~resulting in lower energies in \eref{eq:XX-Ham} when increasing $D$ --- more details below). We also find that it is often required that the interior-point optimizations to be performed inside global minimum finder 
routines, setting a small enough step size tolerance, and fixing the desired constraint tolerance; this is because a caveat of our method is that the optimizer stops prematurely
in numerous possible local minima or may converge to unphysical solutions.

Overall, the required steps for our iterative algorithm %of explicit interior-point optimizations for the iMPS 
can be summarized as follows.
\begin{enumerate}
 \item (Initialization) For a given $D$-value, initialize by constructing a set of $\{A\}$ to 
 form a trial wave function, as in \eref{eq:MPS-WF}, by 
 generating elements \emph{randomly}. Then set \eref{eq:PoissonianCollectionAnsatz} as the constraints 
 for the optimizer, wherever it is \emph{not} possible to apply them in an in-built manner.
 \item (Interior-point iteration) Run one iteration of the interior-point algorithm (or similar) that 
 minimizes an expectation value of the form \eref{eq:C-final} strictly subjected to the constraints of
 \eref{eq:PoissonianCollectionAnsatz} leading to a final set of optimized matrices, $\{A^*\}$. 
 %The run-time complexity of the 
 %finite difference part scales at worst as $O(D^3)$; additionally,
 The most costly operation in each iteration
 comes from either applying $\bar{\mathcal{T}}^{-1}$ on vectors/matrices to evaluate the objective function or the ``direct'' step of the interior-point part scaling at $O(D^3)$. If the function evaluation can be performed through some discrete method efficiently, e.g.~banded LU-decomposition, this part can be done at best with the complexity of $O(D^2)$; otherwise, the iterative methods such as GMRES, 
 also forming and approximately solving a linear system of equations, are readily available and reliable, and and have the complexity of $O(D^2)$ again. Therefore, the dominant cost is always at worst 
 $O(D^{3})$ per this iteration step.
 \item (Checking a stopping criteria) Stop the iterations if the \emph{step size} of the 
 interior-point algorithm drops below a desired tolerance, otherwise return to the previous step.
\end{enumerate}
Practically, the final step above can be replaced by checking for an energy convergence criterion,  
e.g.~$|\frak{e}^*-\frak{e}| < \rm{tol}$, or other robust criteria. Furthermore, although 
no multiple sweeping of the unit-cell sites is necessary above to reach to a fixed point
(unlike iDMRG), the real cost is that the second step must be usually repeated for the 
total number of iterations of $O(D)$ (or worse because of the repetitions required by the global minimum finder) to reach an acceptable accuracy; %in practice; 
this is while, in every
step of iDMRG and VUMPS algorithms, the optimizers often only requires few iterations to diagonalize Hamiltonians in the %MPO 
relevant forms. 

We, therefore, estimate that the global runtime of an interior-point iMPS instances scale roughly as $O(D^4)$. In practice, for the examples given below, we observed an almost $O(D^4)$-scaling and that the large-dimension calculations with $D\sim100$ were completed in the span of few hours using modern-day high-performance processors. This is while the runtime of iDMRG can in practice scales as $O(D^3)$ (usually taking minutes to be completed for $D\sim100$) when it is possible to attain a desirably small truncation error with a fixed number of sweeps. More importantly, iDMRG can heavily benefit from gradually increasing $D$ (decreasing truncation error) by initializing calculations from well-converged wave functions with a smaller $D$. This is currently \emph{not} implemented for interior-point iMPS and would lead to even slower performances comparably.  
However, there are still few reasons to choose interior-point iMPS over 
well-established methods, which we discuss below.   

Here, we
explicitly summarize the connections, differences, 
and potential advantages of the presented method in comparison to iDMRG 
and VUMPS solvers concerning our specific purpose.
\begin{itemize}
 \item (No explicit orthogonalization)
 As in VUMPS, the interior-point iMPS does not require an explicit orthogonalization of the 
 tensor network if enough accuracy is reached in satisfying the constraints (although, we 
 suspect a badly-converged interior-point iMPS may benefit from some explicit orthogonalization 
 routines). This is while 
 iDMRG requires an extra step of explicit orthogonalization after sweeps~\citep{McCulloch2008,McCulloch2008}. 
 %However, 
 %interior-point iMPS by structure always start and retain working with an
 %tensor network chosen to be trivially orthogonalized. 
%  Furthermore, both iDMRG 
%  and VUMPS solvers (usually scaling slightly better than interior-point iMPS as $O(D^3)$ per iteration) 
%  involve truncating the eigenspace of the density matrix and 
%  diagonalizing the Hamiltonian in its MPO representation, which are \emph{not} required in 
%  our method. Therefore, our algorithm stands as a proof-of-principle that tensor network optimizations 
%  can be efficiently performed free from any truncation step; this could 
%  offer more simplicity in the implementation of the algorithm and become useful for pedagogical purposes.
 %(No truncation and MPOs involved) The presented algorithm always work with a
 %translation-invariant iMPS satisfying \eref{eq:PoissonianCollectionAnsatz}, while 
 %keep exploiting all the eigenvalues of $\rho$ to estimate an expectation value. 
 %Meanwhile, iDMRG %and VUMPS 
 %truncate the eigenspace of the density matrix in a number of 
 %sweeps, renormalize it, and find the expectation values by quickly diagonalizing the Hamiltonian
 %in its MPO representation. The latter can be, in general, significantly more efficient, but 
 %writing down and efficiently working with the MPO forms of infinite-range Hamiltonian terms
 %require some careful attention~\citep{McCullochComm}.
 %(Note while VUMPS 
 %is, in fact, more closely-related to our algorithm, it still relies on diagonalization of %effective Hamiltonians
 %in their MPO forms.)
 \item (Simple implementation of symmetry constraints including \eref{rhossAs} and more general cases) Firstly, the direct built-in implementation of 
\eref{rhossAs} is unique to our work,
 and numerical investigations suggest it is effective for stable iterative
 optimizations, which leads to well-converge and reliable $A$-matrices. However, note \eref{rhossAs} is still valid and should exist implicitly for 
 iDMRG and VUMPS formalism as well (as the orthonormality %\eref{eq:ExplicitLeftOrth} 
 and fixed-point equations get immediately or eventually satisfied there too). This is while
 %More generally speaking,
 more realistic
 infinite-dimensional symmetric systems often
 contains symmetry constraints of a modified form of $A^{[j]}_{mn}\neq0~\text{iff}~f(m) = j + f(n)$, where $f$ is a nonlinear analytic function in general (see for example the discussions on tensor network simulations of multi-mode laser cavities in Ref.~\cite{BakerInPrep}). While it might be not straightforward to implement such constraints in iDMRG and VUMPS in a built-in manner, for the interior-point iMPS, $f(m) = j + f(n)$ means a modified form of \eref{eq:A-forms} (and \eref{rhossAs}), which can be immediately implemented as a nonlinear constraint for the nonlinear optimizations.  
 \item (Geometric series infinite-sum relations) Only the VUMPS algorithm is known to exploit geometric series infinite-sum relations for reduced eigen-space of $\cal{T}$-type matrices, equivalent to the ones used in \eref{eq:C-final}, to efficiently calculate some bounding eigenvectors. In fact, 
 similar to Ref.~\cite{ZaunerStauber2018}, the essence of the present work
 is calculating the full 
 energy-per-site expectation values employing direct optimizations of such geometric series infinite-sum terms for the 
infinite-range-interacting systems.
\end{itemize}

% \section{Summarizing the connections to existing tensor network solvers}
% 
% It is important to clarify that our constrained iMPS optimization approach is similar to iMPS and
% VUMPS in the sense of ... Furthermore, our independently-developed method is almost equivalent to 
% VUMPS except the following subtle details: ... These subtle differences lead to the 
% clear advantage of our approach: ... 

\section{Some Energy Results}
In this section, we present the results of a proof-of-concept and relatively small scale interior-point iMPS simulation for 
the Hamiltonian in \eref{eq:XX-Ham}.
We systematically followed the steps discussed above to 
find the global minimum of the energy-per-site for $H_{XX}$, i.e.~employing
Eq.~\eqref{eq:C-final}, for some initially randomized $A$-matrices' elements, subjected to \eref{eq:PoissonianCollectionAnsatz}, while keeping 
small enough step size and fixing all other %constraint
interior-point tolerances to $1e-13$. 
%for all $D$-values.
%the function tolerance of $10^{-6}$, constraints tolerance of $10^{-14}$, and step tolerance of $10^{-8}$.
We have performed the calculations on a %modern-day
high-performance computing system exploiting parallelization --- for even larger scale interior-point iMPS simulations performed using similar %state-of-the-art 
parallelization methods on supercomputers see Ref.~\citep{BakerInPrep}, where 
a continuous-beam laser is consistently modelled.
Notice also that the memory cost of our optimizations remained 
highly manageable as we always exploit efficient methods to save and manipulate matrices as sparse-type inputs.
Overall, these allowed us to efficiently find well-converged iMPS ground states of $H_{XX}$ for bond dimensions up to $D_{\max}=200$, i.e.~optimizing maximally $597$ free parameters.

%%%%%%%%%%%%%%%%%%%%%%% FIG 1 %%%%%%%%%%%%%%%%%%%%%%%%
\begin{figure}[bth]
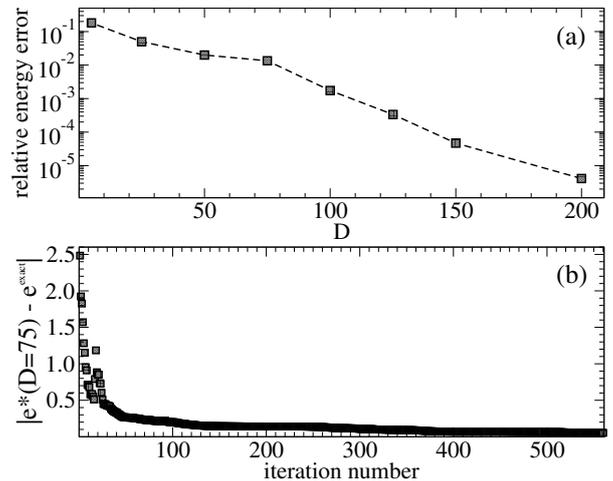

  \includegraphics[width=0.91\linewidth]{{{Fig1}}}
  \caption{(a) The relative error, $|\frac{e^*(D)-\frak{e}^{\rm exact}}{\frak{e}^{\rm exact}}|$, in calculating the ground-state energies per site of $H_{XX}$, \eref{eq:XX-Ham},
  from the proposed interior-point iMPS algorithm, $\frak{e}^*(D)$, and the exact lower
  energy-per-site bound, $\frak{e}^{\rm exact}=-4$, based on the argument presented in the text. Error bars 
  are smaller than the symbol size and the dashed line is provided only as a guide for the eye. (b) The iMPS ground state energies for $H_{XX}$ versus iteration number 
  for a selected series of interior point calculations with $D=75$ (notice each iteration of the 
  interior-point algorithm itself typically contains hundreds of function evaluations).
  \label{fig:energies}}
\end{figure}
%%%%%%%%%%%%%%%%%%%%%%%%%%%%%%%%%%%%%%%%%%%%%%%%%%%%%%

The energies per site, $\frak{e}^*$, for 
the optimized interior-point iMPS ground states of $H_{XX}$ and selected bond dimensions are presented in \fref{fig:energies}(a). There, we report the energy difference with respect to a 
reference/benchmark ground-state energy per site, $\frak{e}^{\rm exact}=-4$, which   
we argue is achievable (e.g.~through an iMPS representation with $D\rightarrow\infty$) and analytically 
show in the following that is an \emph{exact} lower bound on the energies of the Hamiltonian~\citep{McCullochComm}. 
The Hamiltonian can be written as 
$H_{XX} = (\hat{S}^+_{\rm total}\hat{S}^-_{\rm total} + \text{h.c.}) - \sum_{i} (\hat{S}^+_i \hat{S}^-_i + \text{h.c.})$. 
The first term in this form is completely positive and its expectation value is lower bounded by zero. Therefore,
if we find the state that maximizes the expectation value of $\sum_{i} (\hat{S}^+_i \hat{S}^-_i)$ 
and essentially satisfies $\hat{S}^+_{\rm total}\ket{\text{ground}}=0$ and $\hat{S}^-_{\rm total}\ket{\text{ground}}=0$, in principle, 
it would set the lower bound on the energy per site of $H_{XX}$.
Consequently, the true paramagnetic ground state can be perturbatively (ignoring the normalization) written as
\begin{align}
\ket{\text{ground}}_{\rm exact}\propto&\ket{\cdots,0,0,0,\cdots}+\notag\\
                    \sum_{\rm permut.~of~\pm1~and~0s} &\ket{\cdots,0,1,\cdots,0,\cdots,-1,0,\cdots} + \notag\\
                    \sum_{\rm permut.~of~\pm1~and~0s} 
                     &\ket{\cdots,0,1,1,\cdots,0,\cdots,-1,-1,0,\cdots} \notag\\
                     &+ \cdots,
\label{eq:exact}                     
\end{align}
which guarantees that the expectation values of $\hat{S}^+_{\rm total}$ and $\hat{S}^-_{\rm total}$ vanish 
and leads to maximum energy per site value of $4$ for 
the expectation value of $\sum_{i} (\hat{S}^+_i \hat{S}^-_i+ \text{h.c.})$ for spin-$1$ particles.
% from single-site~\citep{Hubig2015} fDMRG calculations on a 100-site lattice with 
% a fixed and relatively large $250$ number of states. For the fDMRG, 
% we employ built-in U(1)-symmetries and the same programming environment on the same personal computer
% on which the interior-point optimizations were performed.
% %, shown as $\frak{e}^{\rm fDMRG}_{\{L=100,D=200\}}$. More presicely speaking, $\frak{e}^{\rm ED}_{\infty}$ is the extrapolated ground-state energy per site found from (pseudo-exact) Lanczos diagonalizations of the finite-size XX Hamiltonian with only built-in U(1) 
% %symmetry~\citep{Sandvik2010} (to match U(1)-symmetry settings in tensor network calculations).
% %For the ED calculations, we found the paramagnetic ground states of $H_{XX}$ on rings of sizes $L=4,6,8,10$ sites and then %extrapolated the energy toward $L\rightarrow\infty$ using conventional least square fitting.
% (Notice that fDMRG can provide far better benchmark energies in comparison to extrapolating the energies from some small-size
% ED calculations.) It is important to note that although we have only presented few interior-point iMPS energy 
% points for $H_{XX}$  but 
% their order of magnitude remain unchanged as $D$ increases; our investigations showed that it is 
% also possible to systematically extrapolate ED and fDMRG ground state energies per site to a finite value as system size 
% increases (not presented here). While no exact lower energy bound is known for the infinite-range infinite-size $H_{XX}$, the discussed results
% suggest a true ground state exist.

It is clear from \fref{fig:energies}(a) that the iMPS wave function's energy per site consistently decreases
toward the true ground state as $D$ increases as desired. (We have also empirically confirmed that the correlation functions of type $\la \hat{S}^+_r \hat{S}^-_0 \ra$ decay, indeed, exponentially with $r$ as it is expected from such iMPS ground states.) In addition, \fref{fig:energies}(b) demonstrates that such interior-point iMPS 
optimization is, after doing enough number of initial iterations, systematically converging toward a true energy minimum as more iterations are 
performed for this exemplary series of calculations 
with $D\!=\!75$.
% In other words, the 
% interior-point iMPS method seems to successfully produce paramagnetic ground states with energies increasingly 
% closer to 
% $\frak{e}^{\rm exact}$ as the bond dimension
% increases. 
%while the energy difference (as a measure of the energy error) seems to reduce polynomially versus $D^{-1}$, as expected. %, and eventually becoming even better than not-so-precise ED results;
% Notice though the iMPS wave functions, even with the smallest bond dimension of $D=20$, \emph{better} represent (have lower energies) than the best finite-size ED result for $L=12$, which is about the ring length that modern-day 
% personal computers can efficiently handle, as demonstrated in the figure.
We argue these results provide strong support for the effectiveness and functionality of 
the interior-point iMPS algorithm.
%We demonstrate the element-wise structure of iMPS $A$-matrices for the representative value of 
%$D=100$ in Fig ...(a) 
(Let us also note that some patterns in the structure of the $A$-matrices' elements may be
physically irrelevant in practice; in particular, while the ratios between magnitudes is physically important,
some other values have been observed to become insignificantly small due to the fact that the algorithm has to stop after 
some finite number of iterations~\citep{BakerInPrep}.) 
%Finally, from both iMPS and ED results, we observe that 
%the stabilized ground state is indeed a paramagnet. 

In the end, we briefly review the physical consequences of the results in \fref{fig:energies}(a). The fact that the iMPS is converging closer and closer to the analytical state in \eref{eq:exact} with energy $\frak{e}^{\rm exact}$ , which corroborates that the perturbatively presented wave function is indeed the true ground state. Now, it can be also easily verified (using a similar argument as above) that the ground states of one-dimensional global-range antiferromagnetic XX Heisenberg models, as in \eref{eq:XX-Ham}, with $J>0$ and few sites are some paramagnets having superpositions of permuting $\{-1,0,1\}$ patterns. In fact, here, we analytically and numerically established that for the infinite-range case of \eref{eq:XX-Ham} the true ground state is again a similar paramagnet possessing the specific permuting patterns shown in \eref{eq:exact}. This is while, in a distinct way, the paramagnetic phase of the nearest-neighbor XX model lives on a critical XY-Haldane point of XYZ Heisenberg Hamiltonian with no infinite-order phase transition~\cite{Malvezzi2016}. Our results provide a framework for future works to study infinite-dimensional variants of XYZ Hamiltonian to investigate possible existence/absence of Kosterlitz-Thouless-type transitions.

\section{Conclusion}
We have presented an efficient iterative tensor network approach to systematically find the ground states
of infinite-dimensional spin Hamiltonians based on explicit constrained optimizations of their iMPS description.
We exemplified how to greatly reduce the number of free parameters in the optimizations
by employing built-in symmetries for the iMPS ansatz.
Previously, the phase diagram of such Hamiltonians have been only studied in the thermodynamic limit 
for a number of exactly-solvable cases 
%or using not-so-accurate mean field theory approaches 
to our knowledge.
We therefore offered a new tensor network algorithm specialized for scrutinizing extreme non-locality of 
infinite-size systems exhibiting infinite-range interactions. 

The presented algorithm
sits next to the slightly more efficient generic tensor-network solvers of iDMRG and VUMPS, which
can be employed to find ground states of non-decaying Hamiltonians in the thermodynamic limit as well. 
In other words, our results demonstrate that one can also derive such ground states by directly optimizing iMPS operators' elements, while 
relying on \emph{no} density matrix truncation and extra explicit orthogonalization steps (as also previously 
claimed in Ref.~\citep{Ostlund1995}); 
%neither any MPO representation, 
this could offer a simpler structure for implementation of the symmetry constraints and usefulness for pedagogical purposes.   
We expect that the phase diagrams of a wide range of infinite-range models of experimental or foundational importance 
(most notably variants of quantum Dicke-Bose-Hubbard-type models as in cavity-mediated bosonic experiments and infinite-dimensional XYZ Heisenberg Hamiltonians) can be now elucidated using interior-point iMPS to guide the direction of the future 
experiments. 
%Nevertheless, the following question remains to be further investigated: 
%how this constrained optimization of iMPS compete against state-of-the-art 
%iDMRG and VUMPS solvers to find the ground states of a typical local Hamiltonian? Our results suggest 
%that our algorithm is somewhat
%slower to converge to the true ground state and harder to control, but improvements might be possible.

Perhaps, the main physical application for the presented algorithm is to provide an optimization 
approach for infinite-dimensional effective Heisenberg Hamiltonians emulating idealistic multi-mode laser models~\citep{BakerInPrep}: in these systems, the symmetry constraints finds complicated algebraic forms (in comparison to, e.g.,  $\{ A^{[j]}_{m,n} \neq 0 
~\text{iff}~ m+j=n \}$ used above) and may become incompatible with existing symmetric iDMRG and VUMPS routines. Nevertheless, the brute-force interior-point iMPS can always handle closed forms of symmetry equations as built-in nonlinear constraints.

\begin{acknowledgments}
  SNS is indebted to Howard Wiseman and Ian McCulloch for some original ideas, useful comments, and inspiring discussions. SNS is also grateful for the helpful comments and encouragement to complete this work from Tim Gould and Joan Vaccaro.  
  This research was funded by the Australian Research Council Discovery Project DP170101734. SNS acknowledges the traditional owners of the land on which this work was undertaken at Griffith University, the Yuggera people.
\end{acknowledgments}

\bibliographystyle{apsrev4-1}

\cleardoublepage
\end{document}